\documentclass{llncs}
\usepackage[pdftex]{graphicx}  
\usepackage[utf8]{inputenc}
\usepackage{fourier} 
\usepackage{array}
\usepackage{makecell}
\usepackage{hyperref}

\begin{document}

\title{Mining Open Government Data Used in Scientific Research}
\titlerunning{OGD Use}  
%
\author{An Yan\inst{1} \and Nicholas Weber \inst{1}}
%
\authorrunning{Yan and Weber} 
%
\tocauthor{An Yan, Nic Weber}
\institute{The Information School, University of Washington\\
\email{yanan15@uw.edu, nmweber@uw.edu}}

\maketitle              

\begin{abstract}
In the following paper, we describe results from mining citations, mentions, and links to open government data (OGD) in peer-reviewed literature. We inductively develop a method for categorizing how OGD are used by different research communities, and provide descriptive statistics about the publication years, publication outlets, and OGD sources. Our results demonstrate that, 1. The use of OGD in research is steadily increasing from 2009 to 2016;  2. Researchers use OGD from 96 different open government data portals, with data.gov.uk and data.gov being the most frequent sources; and, 3. Contrary to previous findings, we provide evidence suggesting that OGD from developing nations, notably India and Kenya, are being frequently used to fuel scientific discoveries. The findings of this paper contribute to ongoing research agendas aimed at tracking the impact of open government data initiatives, and provides an initial description of how open government data are valuable to diverse scientific research communities.

\end{abstract}
\section{Introduction}
The release of public sector information (PSI) has traditionally been motivated by democratic ideals related to representative governance, accountability, and transparency. Over the last decade, governments around the world have greatly expand the scope of PSI accessibility through their participation in e-democracy and e- government initiatives \cite{Jaeger}. Most notably, there is a global movement towards publishing openly licensed, machine-readable data, known as Open Government Data (OGD), from transportation, budgeting, agricultural, and public health agencies \cite{Davies}.\\ 
\indent While much of the public support for open government data remains focused on increasing government transparency, public officials have also begun to recognize the value of open data for the private sector \cite{Zuiderwijk}. Thus, much of the existing research into open government data initiatives has focused on two questions: First, how does open government data affect government accountability and transparency ? \cite{Jaeger}; and, Second, in opening up government data to entrepreneurs, what economic impact does open government data create? \cite{Gruen}. In this paper, we seek to expand current efforts in tracking the impact and value of open data initiatives by exploring the use of OGD in scientific research. By mining citations, mentions, and links to open government data found in a broad collection of peer-reviewed literature, we present empirical evidence about which scientific research communities are using OGD, and for what purposes. In the following section, we situate our work amongst previous studies of OGD users, and then present the research design for this study. 

\section{Related Work}
Although OGD has sparked a growing level of interest among researchers, there is little empirical work that focuses explicitly on how these resources are used, in practice, by the public \cite{Zuiderwijk_2015}. Notable exceptions include, Bright et al. \cite{Bright} who analyzed downloads of data hosted by data.gov.uk. They found that a majority of OGD has never been accessed, and the most frequent use of OGD was for commercial purposes. Similarly, Young and Yan \cite{Young}  examined the challenges and expectations of civic hackers using open government in developing new technologies. They found that this community expressed a desire for higher quality data, and that issues related to the functionality of an open data portal greatly limited OGD use.\\
\indent There are few studies that have examined OGD and its potential use in scientific research. Safarov et al. \cite{Safarov} conducted a systematic literature review examining the utilization of open government data through peer-reviewed publications. These authors categorize open government data use based on the type, condition, effect, and users of open data. They argue that OGD could allow researchers to form new analysis based on OGD, but note that there is little evidence suggesting that scientific communities are interested in this data source. In one of the few targeted studies looking into how open government data is actually being used in scientific research, Martin et al. \cite{Martin} examined the requirements for and potential uses of the New York State's open health data portal. Collecting data through surveys and focus groups, they argue that obstacles to OGD use in research include low awareness of data availability and limited engagement with government data producers.\\
\indent This paper seeks to better understand the use of open government data for scientific research by answering the following research questions:
\begin{itemize}
\item How are OGD used in academic research? 
\item What sources of OGD do researchers use? 
\item What fields of academic research are using OGD?
\end{itemize}
In the following sections, we describe the methods used for identifying and analyzing citations, mentions, and links to open government data in peer reviewed literature published between January 2009 and July 2017. We summarize our findings through descriptive statistics, and preliminary observations about how open data are used by different scientific research communities.

\section{Methods}
We developed a systematic review protocol to identify, select, and categorize published literature where open government data was explicitly used as a research input. By following a systematic review protocol this study is replicable, in that our results can be reproduced by following the steps outlined below, and exhaustive, as our search for relevant literature makes use of all potential sources currently indexed by three major publication databases \cite{Fecher}. In the following sections, we describe the systematic protocol used for identifying and selecting relevant literature, and the methods used for mining this literature for evidence of OGD use.\\ 

\noindent \textbf{Search Terms:} We first obtained a list of open government data portals from data.gov\footnote{https://www.data.gov/open-gov/}. We chose to remove portals from this list that were not specific to one particular government or agency (e.g. The World Bank), and we also added well-respected international OGD sources that were absent from the initial list (e.g. Taiwan - data.gov.tw). In total, there were 302 unique URLs representing sources where researchers might obtain open government data. Table \ref{tab:search_terms} provides a summary of the types of OGD portals included in this study.\\

\begin{table}
\centering
\caption{Summary of OGD portals used as search terms}\label{tab:search_terms}
\begin{tabular}{p{3cm}p{0.8cm}lp{4cm}}
\hline\noalign{\smallskip}
Type                   & Count & Examples                    \\
\noalign{\smallskip}
\hline
\noalign{\smallskip}
International Regional & 155   & Victoria (http://www.data.vic.gov.au/)                \\
International Country  & 54    & Australia (http://data.gov.au/)                       \\
US City or County      & 46    & Seattle (http://data.seattle.gov/)                    \\
US State               & 38    & Hawaii (http://data.hawaii.gov/)                      \\
Other State Related    & 8     & NY Department of Health(https://health.data.ny.gov/)  \\
US                     & 1     & Data.gov (https://www.data.gov/)           \\          
\hline
\end{tabular}
\end{table}

\noindent \textbf{Article Retrieval:} We constructed a query combining the 302 data portal URLs described above, and used this query to search the Scopus, Springer, and IEEE databases in July, 2017. Queries were limited, by date, to January 2009 to July 2017 in order to align with the establishment of major open data initiatives, such as the launch of data.gov in 2009. In total, 2486 articles were identified. 

Next, we removed duplicate publications that appeared in multiple databases.  We then further refined the study sample by selecting only peer-reviewed full-text journal articles or conference papers (i.e. we did not include  abstracts, chapters, or books). We only selected publications that were written in English, had full-text accessible to our research team, and in which OGD was actually used by the authors, and not simply mentioned as being a relevant topic or phenomena of interest. To determine which papers should remain in the sample using this selection criteria, the two authors of this paper individually examined 170 articles selected randomly from the initial sample of 2039 publications. Cohen’s Kappa was calculated to determine selection agreement, and Landis and Koch’s guideline was used to interpret the kappa statistics \cite{Landis}. The authors achieved substantial agreement for both inclusion and exclusion of relevant papers. The first author then applied the selection criteria to the remaining papers in the initial sample. This process yielded a study sample of 1229 papers (See Figure 1).\\

\begin{figure}
\centering
\label{fig:flow}
\includegraphics[width=0.8\textwidth]{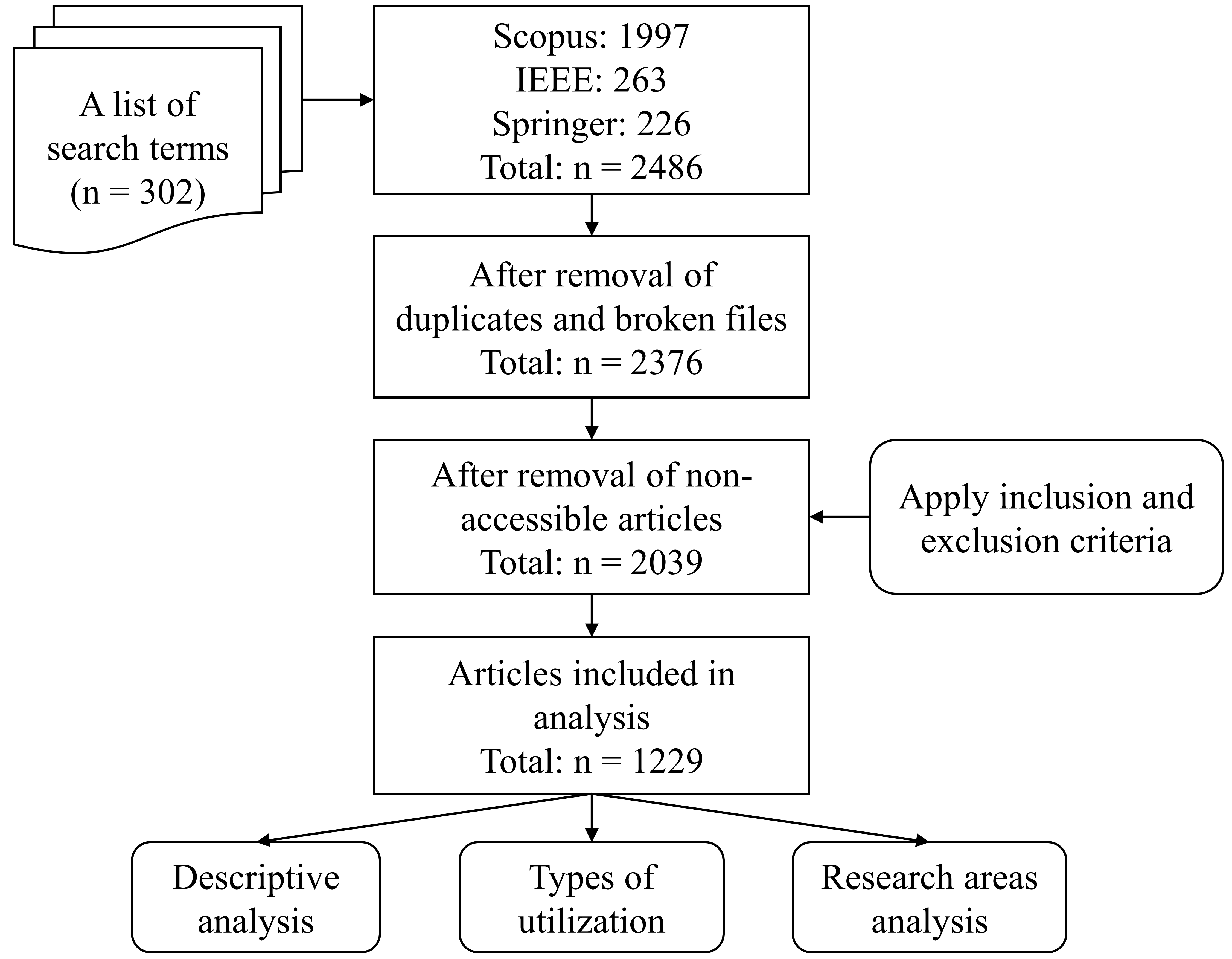}
\caption{The process of literature mining}
\end{figure}

\noindent \textbf{Determination of Usage Types}: Using the study sample (n=1229), the two authors read through the full text of a randomly selected subset of 63 papers. While reading the two authors inductively developed separate codebooks to categorize different ways that the publications described using OGD. The authors then met, combined their codebooks, and recoded an additional 200 articles. Seven usage types emerged from this inductive coding: OGD were used 1) As the main data source for new analysis, 2) As ancillary data source for new analysis, 3) For result evaluation, 4) To demonstrate the effectiveness of a proposed new method , 5) To develop information services or new platforms, 6) To create a combined or composite dataset, and 7) To provide a broader context for a study's subject or topic. We noted that the categories were not mutually exclusive; an author may use multiple OGD sources in a single publication, or may use a single OGD source for multiple purposes in the same publication. To ensure our shared understanding of the merged codebook, Cohen’s Kappa was calculated in a final round of coding (n=50 articles). Nearly perfect agreement was achieved for usage types 1), 2), 5), 6), and 7), and substantial agreement was achieved for usage types 3) and 4). The first author then coded the rest of the samples based on the mutual understanding of the codebook.\\

\noindent \textbf{Additional Analysis of Sample}: We analyzed publications in our sample based on publication year, in order to understand the trend of OGD usage over time, as well as the source of OGD, in order to understand which data portals were most popular. We also analyzed the publication outlets (e.g. journal, conference, etc) to gain an understanding of how various research fields engage with OGD. We associated each outlet with its research area using Scopus source title list \footnote{https://www.elsevier.com/solutions/scopus/content} which provided  information of each publication outlet indexed by Scopus. We discuss these results below.\\

\section{Results}
\begin{figure}
\centering
\includegraphics[width=0.6\textwidth]{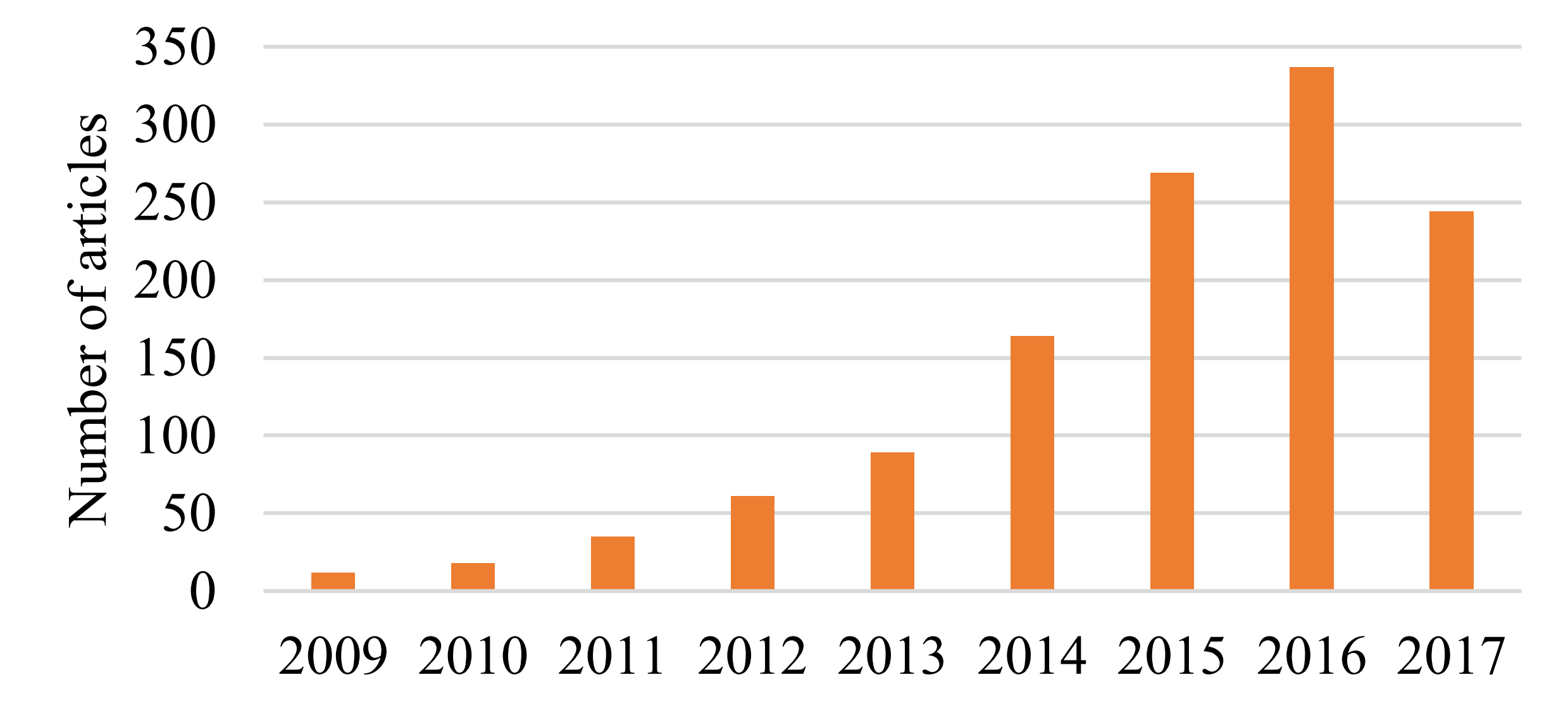}
\caption{Number of articles using OGD by year}\label{fig:year}
\end{figure}

\textbf{Descriptive Analysis:} Figure \ref{fig:year} shows the number of articles that used OGD from January 2009 to July 2017. There is an upward trend of publications using OGD over time, implying the increasing attention to OGD in scientific research. This is consistent with previous findings that as open data initiatives mature, portals are more frequently used by diverse stakeholders\cite{Thorsby}.\\ 

\begin{figure}
\centering

\includegraphics[width=1.0\textwidth]{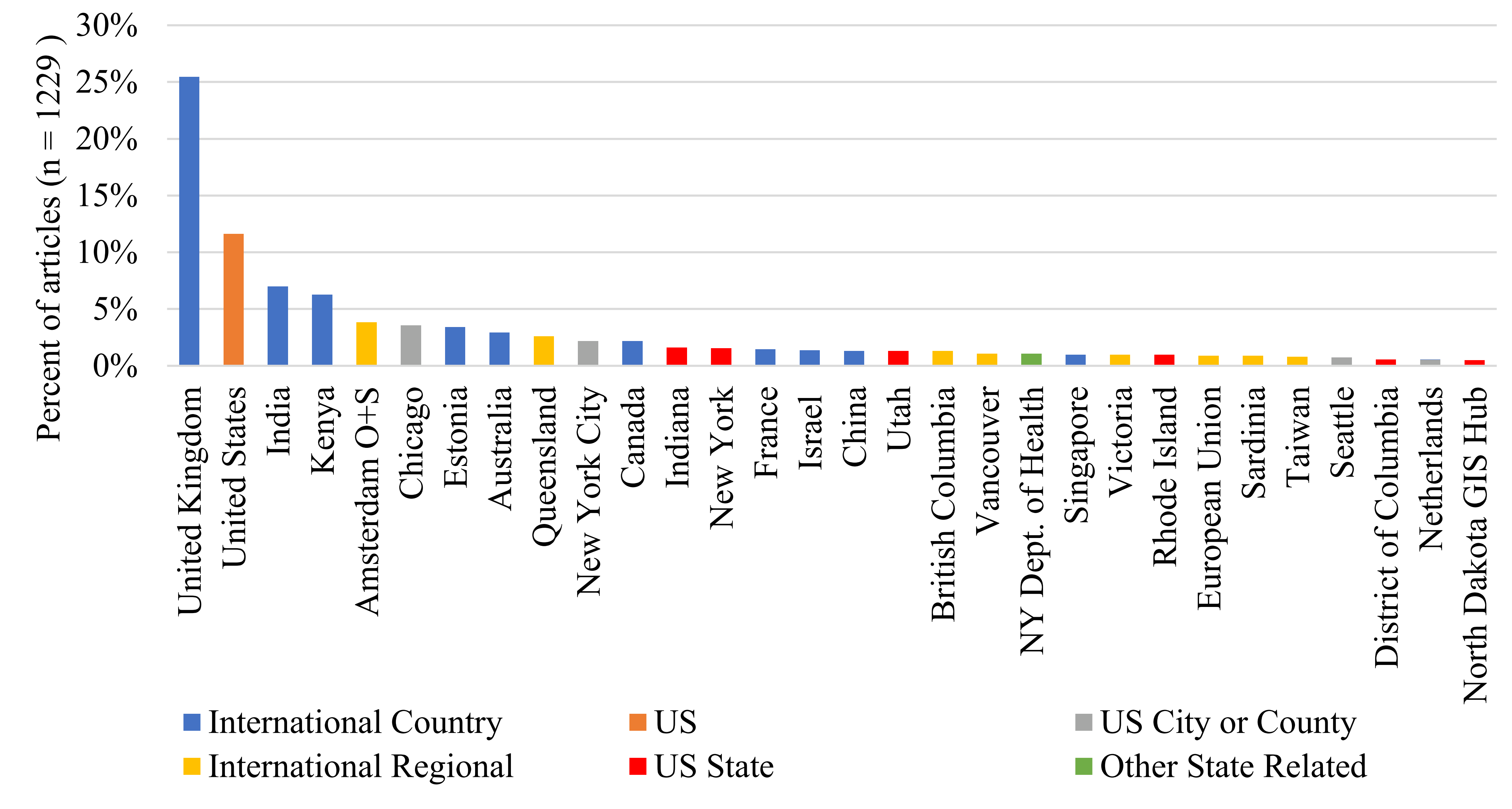}
\caption{Top 30 sources of OGD used in scientific research}\label{fig:sources}
\end{figure}

The sources of OGD research are provided in Figure \ref{fig:sources}. Data from 96 different Open Government Data Portals were used by publications in our sample. Notably, the United Kingdom's OGD portal (data.gov.uk) was used by 25.5\% of all papers in our sample. This is more than twice the number of articles (11.6\%) that used the United States OGD portal (data.gov). It is not surprising that national portals are more frequently used than state, province, city or county portals due to the fact that both the UK and USA portals have more data available. National OGD portals, such as data.gov, also harvest data from sub-national portals at the city and state level. However, it is surprising that Chicago is ranked as the sixth most frequent source of OGD (3.6 \%). Chicago is the most popular OGD portal among all US city or county level portals. This is also surprising as other major USA cities such as New York City, Seattle, and Boston have more data available for download, and have holdings which are more extensive than those of Chicago \cite{Kassen}. Among international regions, Amsterdam (3.8\%) and Queensland (2.6\%) received the most use. This result is in line with Safarov et al.’s \cite{Safarov} findings that OGD-related studies were most active in developed countries with the Netherlands, the United States, the United Kingdom as the most researched countries. However, OGD in two developing countries, India (7.0\%) and Kenya (6.3\%), notably appear in the top 5 of all sources found in our study. This is a significant finding, as it suggests that although these two nations have relatively new OGD programs, they are a unique and valuable source of data for researchers working on topics related to these countries.\\

\textbf{Research Areas That Use OGD:} We examined the publication outlets for the selected articles in our sample (n=1229). Among the 815 different publication outlets, PloS ONE produced the most (32) articles, followed by BMC Public Health (14), Science of the Total Environment(12), Sustainability (Switzerland) (11), BMJ Open (11), and Scientific Reports (9). Notably, all of these journals, with the exception of Science of the Total Environment, are open-access publications. This suggests that authors who use open data may be more likely to publish in open-access venues, or that these publication outlets have more strict standards about data source citation. However, further research is required to validate this observation.\\ 
\indent We also associated the title of each publication outlet in our sample with a corresponding field of research as defined by the Scopus database. Table 2 shows that Medicine, Environmental Sciences and Social Sciences occupy the top three fields that use OGD in publications, suggesting that OGD is playing the role of advancing both natural science and social science discoveries. Computer Sciences and Engineering also rank high in the list (No. 4 and No. 6), suggesting that OGD is also contributing to technical innovations. Noticeably, OGD was used by all available Scopus research fields, including  Dentistry and Chemistry. We caution that the research area of a publication outlet does not necessarily represent the research areas of a paper published in it, but we do argue that these labels are a strong proximate indicator of the topic of the article. Thus, the breadth and diversity of research fields using OGD demonstrates that these PSI are indeed a valuable source for scientific research.\\

\begin{table}
\centering
\caption{Top 10 research areas that use OGD} \label{tab:fields}
\begin{tabular}{p{6cm}p{1cm}}
\hline\noalign{\smallskip}
Research Areas                               & Count \\
\noalign{\smallskip}
\hline
\noalign{\smallskip}
Medicine                                     & 557   \\
Environmental Science                        & 448   \\
Social Sciences                              & 412   \\
Computer Science                             & 242   \\
Agricultural and Biological Sciences         & 178   \\
Engineering                                  & 145   \\
Earth and Planetary Sciences                 & 144   \\
Biochemistry, Genetics and Molecular Biology & 80    \\
Energy                                       & 76    \\
Business, Management and Accounting          & 45 \\
\hline
\end{tabular}
\end{table}

\begin{figure}
\centering

\includegraphics[width=0.7\textwidth]{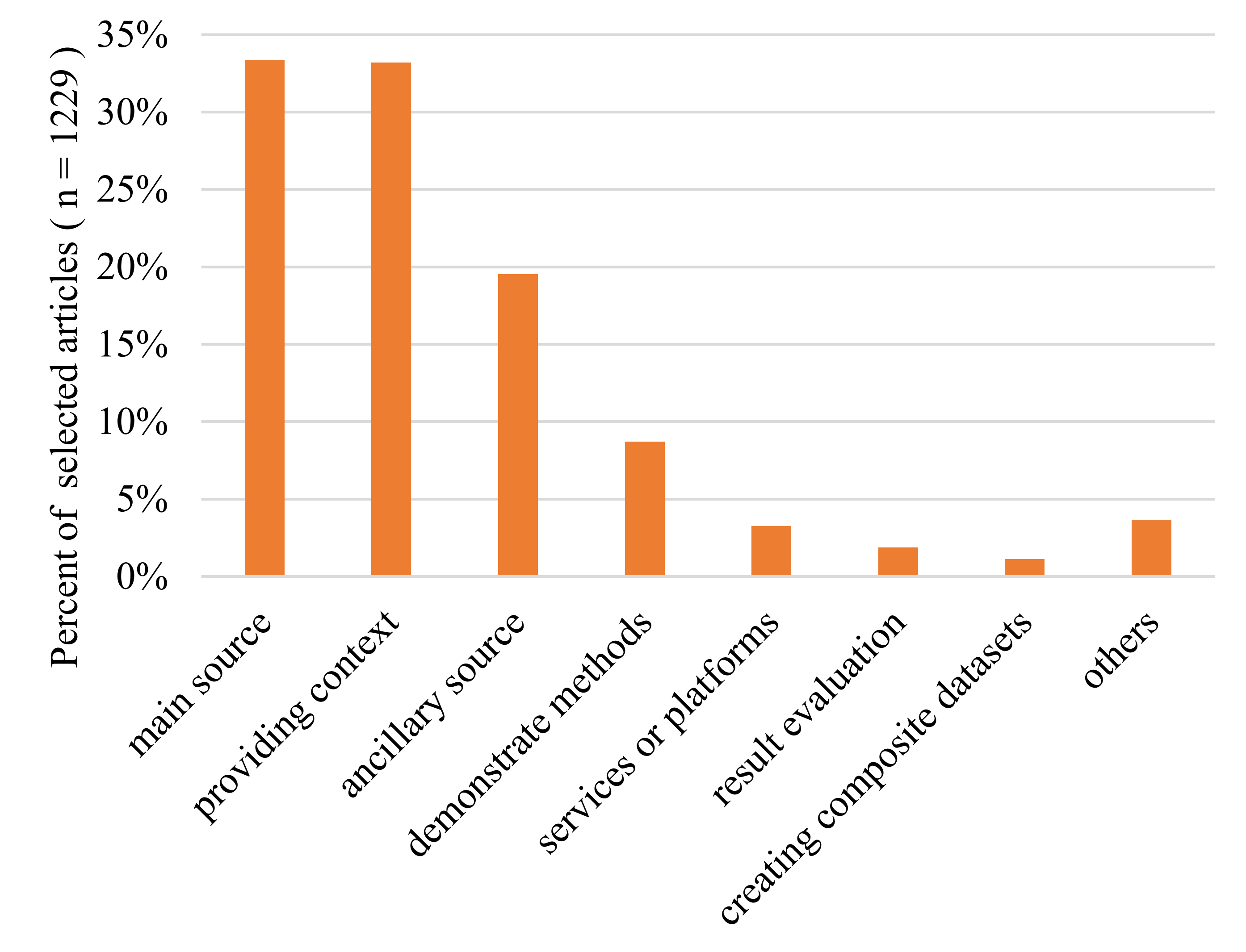}
\caption{Percentage of usage types by article}\label{fig:types}
\end{figure}

\textbf{OGD Usage Types:} Through inductive coding, seven major OGD usage types were identified. OGD were used 1) As the main data source for new analysis, such as being a variable in regression or a parameter in a mathematic model; 2) As ancillary data source for new analysis, such as the base map upon which newly collected data are plotted; 3) For result evaluation, such as to validate or verify a research claim, or to benchmark the quality of an existing dataset; 4) To demonstrate the effectiveness of a proposed new method (e.g., an innovative statistical model); 5) To develop information services or new platforms (e.g., a semantic web browser); 6) To create a combined or composite dataset; and 7) To provide study context, such as describing the motivation for a study. Figure \ref{fig:types} illustrates the distribution of OGD uses from our study sample. Over half of the publications used OGD as data sources for new research: 33.4\% used OGD as main source and 19.5\% as an ancillary source. In our sample, 33.2\% of the articles cited OGD to provide background context for the study, including demographic information about a population of study, or as validation for a chosen sampling method. Less then 5\% of the articles belong to other usage types that were unclear from the text of the article.\\

\noindent \textbf{Limitations:} There are several limitations to this study. First, and most significantly, our search only retrieved articles that explicitly cite, mention, or link to an open government data portal. Many researchers likely use OGD, but do not explicitly document the source. We present these results as preliminary findings about how OGD are used, and discuss future ways in which implicit OGD use could be identified below. Second, we only considered OGD sources retrieved from an initial list of 302 portals. Although we believe this list to be extensive, there are certainly sources of OGD that were not included in our search. Third, we only included three databases in our analysis and most selected articles were retrieved from Scopus, therefore our results relied heavily on the content coverage of the Scopus database. Our categorization of research areas is also constrained by the way that Scopus assigns research areas to each publication outlet. In the future, we hope estimate the impact of OGD on academic research by incorporating journal impact factors, and citation counts of selected articles into our analysis. We also hope to understand  researchers’ practices with OGD by conducting surveys and interviews. By combining the literature mining approach described here with survey and qualitative methods, we believe these results could be improved. Below, we restate the key findings of this initial work.\\ 

\section{Discussion \& Conclusion}
This paper is one of the first empirical examinations of how OGD are being used in scientific research. By mining a total of 1229 research articles that used OGD, we found that 1. The use of OGD in research is steadily increasing from 2009 to 2016; 2. Researchers use OGD from 96 different open government data portals in their publications, with data.gov.uk and data.gov being the most frequent sources; 3. Contrary to previous findings, we provide evidence that OGD from developing nations, notably India and Kenya, are popular amongst OGD users; 4. OGD has been used by nearly all research fields, with Medicine, Environmental Science, and Social Sciences being the most active fields that engage with OGD; and 5. Researchers mainly use OGD as data sources for new analysis and to provide context for a research finding. OGD was also used for testing new methods, providing new services or systems, result evaluation, and creating composite datasets.\\ 

\textbf{Popular OGD Sources:} We discovered in our results that United Kingdom, Kenya, and Chicago ranked high in our sources of OGD. Looking further into the use of these sources, we found out that each of the three portals had one particular type of dataset that had been used most frequently. For example, in the United Kingdom, an "index of deprivation" \footnote{ https://data.gov.uk/dataset/index-of-multiple-deprivation} dataset was used frequently as an ancillary source for Medicine and Public Health research \cite{Evans}. For Kenya, demographic data was often used to describe the study setting for research on poverty in developing countries \cite{Okotto}. For Chicago, crime data \footnote{ https://data.cityofchicago.org/public-safety/crimes-2001-to-present/ijzp-q8t2} was the most used source for social, urban, and computer science studies \cite{Zheng}. The frequent use of a particular dataset from a particular data portal requires further consideration: First, as official data published by a government entity, these datasets may be the only authoritative sources for certain types of information, such as demographic characteristics of a population. This would explain why, for example, the city of Chicago appears high in our rankings of OGD sources. Researchers interested in crime may find this the best, or only source to answer a research question. Second, these key sources of data can help answer questions that are of great local interest, but the remaining datasets published by a government may be of limited use to researchers outside of a particular topic. For example, if we removed poverty data from the Kenya portal we would get substantially different results as to the popularity of developing nations OGD source. But, these results imply that governments should seek to identify highly unique data, and promote their use amongst research communities. In turn, this could spur further engagement with a data portal. Further, governments should consider prioritizing the release of datasets that satisfy the above-mentioned conditions in order to maximize the utilization of their investment in open government data initiatives.\\

\textbf{Research Areas:} Finally, we highlight our finding that nearly all scientific fields are using OGD. Previous studies \cite{Graves} have shown that OGD may be used by researchers to investigate education, policies, health-care, and government activities. But, our results suggest that OGD can also be applied to broad research areas such as chemistry \cite{Croft} and Dentistry \cite{Silva} that seem less likely to use OGD and that OGD has exhibited more potential use for research than reflected in exiting literature.

\section{Acknowledgement}
This research was supported in part by IMLS grant \# RE-40-16-0015-16. Supporting data and in-depth explanation of the methods used in this study can be found at \url{https://github.com/OpenDataLiteracy/iConference_2018}

%

\end{document}